\documentclass[aps,prl,twocolumn,showpacs,superscriptaddress]{revtex4-2} 

\usepackage{graphicx}  % needed for figures
\usepackage{dcolumn}   % needed for some tables
\usepackage{bm}        % for math
\usepackage{amssymb}   % for math
\usepackage{xcolor}    % for math
\usepackage{lipsum}
\usepackage{wrapfig}

\begin{document}

\title{First measurement of hard exclusive $\pi^- \Delta^{++}$ electroproduction beam-spin asymmetries off the proton}

% author list 03/07/2023

\newcommand*{\ANL}{Argonne National Laboratory, Argonne, Illinois 60439}
\newcommand*{\ANLindex}{1}
\affiliation{\ANL}
\newcommand*{\CSUDH}{California State University, Dominguez Hills, Carson, CA 90747}
\newcommand*{\CSUDHindex}{2}
\affiliation{\CSUDH}
\newcommand*{\CANISIUS}{Canisius College, Buffalo, NY}
\newcommand*{\CANISIUSindex}{3}
\affiliation{\CANISIUS}
\newcommand*{\SACLAY}{IRFU, CEA, Universit\'{e} Paris-Saclay, F-91191 Gif-sur-Yvette, France}
\newcommand*{\SACLAYindex}{4}
\affiliation{\SACLAY}
\newcommand*{\CNU}{Christopher Newport University, Newport News, Virginia 23606}
\newcommand*{\CNUindex}{5}
\affiliation{\CNU}
\newcommand*{\UCONN}{University of Connecticut, Storrs, Connecticut 06269}
\newcommand*{\UCONNindex}{6}
\affiliation{\UCONN}
\newcommand*{\DUKE}{Duke University, Durham, North Carolina 27708-0305}
\newcommand*{\DUKEindex}{7}
\affiliation{\DUKE}
\newcommand*{\DUQUESNE}{Duquesne University, 600 Forbes Avenue, Pittsburgh, PA 15282 }
\newcommand*{\DUQUESNEindex}{8}
\affiliation{\DUQUESNE}
\newcommand*{\FU}{Fairfield University, Fairfield CT 06824}
\newcommand*{\FUindex}{9}
\affiliation{\FU}
\newcommand*{\FERRARAU}{Universita' di Ferrara , 44121 Ferrara, Italy}
\newcommand*{\FERRARAUindex}{10}
\affiliation{\FERRARAU}
\newcommand*{\FIU}{Florida International University, Miami, Florida 33199}
\newcommand*{\FIUindex}{11}
\affiliation{\FIU}
\newcommand*{\FSU}{Florida State University, Tallahassee, Florida 32306}
\newcommand*{\FSUindex}{12}
\affiliation{\FSU}
\newcommand*{\GWUI}{The George Washington University, Washington, DC 20052}
\newcommand*{\GWUIindex}{13}
\affiliation{\GWUI}
\newcommand*{\GSIFFN}{GSI Helmholtzzentrum fur Schwerionenforschung GmbH, D-64291 Darmstadt, Germany}
\newcommand*{\GSIFFNindex}{14}
\affiliation{\GSIFFN}
\newcommand*{\INFNFE}{INFN, Sezione di Ferrara, 44100 Ferrara, Italy}
\newcommand*{\INFNFEindex}{15}
\affiliation{\INFNFE}
\newcommand*{\INFNFR}{INFN, Laboratori Nazionali di Frascati, 00044 Frascati, Italy}
\newcommand*{\INFNFRindex}{16}
\affiliation{\INFNFR}
\newcommand*{\INFNGE}{INFN, Sezione di Genova, 16146 Genova, Italy}
\newcommand*{\INFNGEindex}{17}
\affiliation{\INFNGE}
\newcommand*{\INFNRO}{INFN, Sezione di Roma Tor Vergata, 00133 Rome, Italy}
\newcommand*{\INFNROindex}{18}
\affiliation{\INFNRO}
\newcommand*{\INFNTUR}{INFN, Sezione di Torino, 10125 Torino, Italy}
\newcommand*{\INFNTURindex}{19}
\affiliation{\INFNTUR}
\newcommand*{\INFNPAV}{INFN, Sezione di Pavia, 27100 Pavia, Italy}
\newcommand*{\INFNPAVindex}{20}
\affiliation{\INFNPAV}
\newcommand*{\ORSAY}{Universit'{e} Paris-Saclay, CNRS/IN2P3, IJCLab, 91405 Orsay, France}
\newcommand*{\ORSAYindex}{21}
\affiliation{\ORSAY}
\newcommand*{\Juelich}{Institute fur Kernphysik (Juelich), Juelich, Germany}
\newcommand*{\Juelichindex}{22}
\affiliation{\Juelich}
\newcommand*{\JMU}{James Madison University, Harrisonburg, Virginia 22807}
\newcommand*{\JMUindex}{23}
\affiliation{\JMU}
\newcommand*{\KNU}{Kyungpook National University, Daegu 41566, Republic of Korea}
\newcommand*{\KNUindex}{24}
\affiliation{\KNU}
\newcommand*{\LAMAR}{Lamar University, 4400 MLK Blvd, PO Box 10046, Beaumont, Texas 77710}
\newcommand*{\LAMARindex}{25}
\affiliation{\LAMAR}
\newcommand*{\MIT}{Massachusetts Institute of Technology, Cambridge, Massachusetts  02139-4307}
\newcommand*{\MITindex}{26}
\affiliation{\MIT}
\newcommand*{\MISS}{Mississippi State University, Mississippi State, MS 39762-5167}
\newcommand*{\MISSindex}{27}
\affiliation{\MISS}
\newcommand*{\ITEP}{National Research Centre Kurchatov Institute - ITEP, Moscow, 117259, Russia}
\newcommand*{\ITEPindex}{28}
\affiliation{\ITEP}
\newcommand*{\UNH}{University of New Hampshire, Durham, New Hampshire 03824-3568}
\newcommand*{\UNHindex}{29}
\affiliation{\UNH}
\newcommand*{\NMSU}{New Mexico State University, PO Box 30001, Las Cruces, NM 88003, USA}
\newcommand*{\NMSUindex}{30}
\affiliation{\NMSU}
\newcommand*{\NSU}{Norfolk State University, Norfolk, Virginia 23504}
\newcommand*{\NSUindex}{31}
\affiliation{\NSU}
\newcommand*{\OHIOU}{Ohio University, Athens, Ohio  45701}
\newcommand*{\OHIOUindex}{32}
\affiliation{\OHIOU}
\newcommand*{\ODU}{Old Dominion University, Norfolk, Virginia 23529}
\newcommand*{\ODUindex}{33}
\affiliation{\ODU}
\newcommand*{\JLUGiessen}{II Physikalisches Institut der Universitaet Giessen, 35392 Giessen, Germany}
\newcommand*{\JLUGiessenindex}{34}
\affiliation{\JLUGiessen}
\newcommand*{\ROMAII}{Universita' di Roma Tor Vergata, 00133 Rome Italy}
\newcommand*{\ROMAIIindex}{35}
\affiliation{\ROMAII}
\newcommand*{\MSU}{Skobeltsyn Institute of Nuclear Physics, Lomonosov Moscow State University, 119234 Moscow, Russia}
\newcommand*{\MSUindex}{36}
\affiliation{\MSU}
\newcommand*{\SCAROLINA}{University of South Carolina, Columbia, South Carolina 29208}
\newcommand*{\SCAROLINAindex}{37}
\affiliation{\SCAROLINA}
\newcommand*{\TEMPLE}{Temple University,  Philadelphia, PA 19122 }
\newcommand*{\TEMPLEindex}{38}
\affiliation{\TEMPLE}
\newcommand*{\JLAB}{Thomas Jefferson National Accelerator Facility, Newport News, Virginia 23606}
\newcommand*{\JLABindex}{39}
\affiliation{\JLAB}
\newcommand*{\UTFSM}{Universidad T\'{e}cnica Federico Santa Mar\'{i}a, Casilla 110-V Valpara\'{i}so, Chile}
\newcommand*{\UTFSMindex}{40}
\affiliation{\UTFSM}
\newcommand*{\INSUBRIA}{Universit\`{a} degli Studi dell'Insubria, 22100 Como, Italy}
\newcommand*{\INSUBRIAindex}{41}
\affiliation{\INSUBRIA}
\newcommand*{\BRESCIA}{Universit`{a} degli Studi di Brescia, 25123 Brescia, Italy}
\newcommand*{\BRESCIAindex}{42}
\affiliation{\BRESCIA}
\newcommand*{\UCR}{University of California Riverside, 900 University Avenue, Riverside, CA 92521, USA}
\newcommand*{\UCRindex}{43}
\affiliation{\UCR}
\newcommand*{\GLASGOW}{University of Glasgow, Glasgow G12 8QQ, United Kingdom}
\newcommand*{\GLASGOWindex}{44}
\affiliation{\GLASGOW}
\newcommand*{\YORK}{University of York, York YO10 5DD, United Kingdom}
\newcommand*{\YORKindex}{45}
\affiliation{\YORK}
\newcommand*{\VIRGINIA}{University of Virginia, Charlottesville, Virginia 22901}
\newcommand*{\VIRGINIAindex}{46}
\affiliation{\VIRGINIA}
\newcommand*{\WM}{College of William and Mary, Williamsburg, Virginia 23187-8795}
\newcommand*{\WMindex}{47}
\affiliation{\WM}
\newcommand*{\YEREVAN}{Yerevan Physics Institute, 375036 Yerevan, Armenia}
\newcommand*{\YEREVANindex}{48}
\affiliation{\YEREVAN}

\newcommand*{\NOWANL}{Argonne National Laboratory, Argonne, Illinois 60439}
\newcommand*{\NOWJLAB}{Thomas Jefferson National Accelerator Facility, Newport News, Virginia 23606}
 %%%%%%%%%%%%%%% END OF Latex Macros for institute addresses  %%%%%%%%%%%%%%%%%%%%%%%%% 

\author {S.~Diehl} 
\affiliation{\JLUGiessen}
\affiliation{\UCONN}
\author {N.~Trotta} 
\affiliation{\UCONN}
\author {K.~Joo} 
\affiliation{\UCONN}
\author {P.~Achenbach} 
\affiliation{\JLAB}
\author {Z.~Akbar} 
\affiliation{\VIRGINIA}
\affiliation{\FSU}
\author {W.R.~Armstrong} 
\affiliation{\ANL}
\author {H.~Atac} 
\affiliation{\TEMPLE}
\author {H.~Avakian} 
\affiliation{\JLAB}
\author {L.~Baashen} 
\affiliation{\FIU}
\author {N.A.~Baltzell} 
\affiliation{\JLAB}
% ...

\collaboration{The CLAS Collaboration}
\noaffiliation

\begin{abstract}
The polarized cross section ratio $\sigma_{LT'}/\sigma_{0}$ from hard exclusive $\pi^{-} \Delta^{++}$ electroproduction off an unpolarized hydrogen target has been extracted based on beam-spin asymmetry measurements using a 10.2~GeV / 10.6~GeV incident electron beam and the CLAS12 spectrometer at Jefferson Lab. The study, which provides the first observation of this channel in the deep-inelastic regime, focuses on very forward-pion kinematics in the valence regime, and photon virtualities ranging from 1.5 GeV$^{2}$ up to 7 GeV$^{2}$. 
The reaction provides a novel access to the $d$-quark content of the nucleon and to $p \rightarrow \Delta^{++}$ transition generalized parton distributions. A comparison to existing results for hard exclusive $\pi^{+} n$ and $\pi^{0} p$ electroproduction is provided, which shows a clear impact of the excitation mechanism, encoded in transition generalized parton distributions, on the asymmetry.
\end{abstract}


\maketitle

%%%%%%%%%%%%%%%%%%%%%%%%%%%%%%%%%%%%%%%%%%%%%%%%%%%%%%
%\section{\label{sec:Intro} Introduction}

Hard exclusive meson electroproduction provides a powerful tool to study the structure of the nucleon and the underlying reaction dynamics as the process amplitude depends on Generalized Parton Distributions (GPDs) \cite{Rad97-5, Col97-6, Brod94-7}. GPDs enable us to access the three-dimensional (3D) structure of the nucleon by correlating the transverse position and the longitudinal momentum of the quarks and gluons inside the nucleon. For longitudinally polarized virtual photons, the factorization of the process amplitude into a perturbatively calculable hard-scattering part and two soft parts has been proven at large photon virtuality $Q^{2}$, large invariant energy $W$ and fixed Bjorken-x ($x_{B}$) \cite{fact1, Col97-6}. The contribution of transversely polarized virtual photons for which factorization is not explicitly proven, is typically treated as a higher twist-effect in current phenomenological models \cite{previous1}. While the GPD framework is already well established for the study of the three-dimensional structure of the ground state nucleon, theoretical attempts have been made to extend this framework to excited nucleon states \cite{BR05, 13a, 13DVCS}. Within such a framework we can significantly extend our understanding of the strong interaction dynamics underlying the generation of the structure of ground and excited nucleon states as relativistic bound systems of quarks and gluons. For this purpose, another set of GPDs, the so-called transition GPDs, have been introduced. 

For the special case of the $N\to\Delta$ transition, there are in total 16 transition GPDs \cite{13a}. 
The first eight are helicity-non-flip (twist-2) transition GPDs (compared to four chiral even GPDs for the ground state nucleon). The unpolarized twist-2 transition GPDs $G_{1}$ - $G_{4}$ can be related to the Jones-Scadron electromagnetic form factors for the $N \to \Delta$ transition \cite{8,9}, while the polarized transition GPDs $\widetilde{G}_{1}$ - $\widetilde{G}_{4}$ are related to the Adler form factors \cite{9,10,10a}. Similar to ordinary deeply virtual meson production, where the description of the twist-3 sector requires the introduction of four additional transversity GPDs, the description of hard exclusive $N\to N^{*}$ pion production ($e N \to e^\prime N^{*} \pi$) requires the introduction of eight additional helicity flip / transversity transition GPDs ($G_{T_{1}}$ - $G_{T_{8}}$), which describe the impact of the transversely polarized virtual photons on the twist-3 amplitudes \cite{13a}.
Hard exclusive electroproduction of $\pi^{-}\Delta^{++}$ has been theoretically described based on transition GPDs in Ref. \cite{13a}. It has been shown that in total 12 of the 16 transition GPDs contribute to the different observables of exclusive $\pi^{-}\Delta^{++}$ electroproduction. The evaluation of these so far poorly known transition GPDs has been based on symmetry relations in the large $N_{C}$ limit \cite{Frankfurt:1999fp,11,12,13,13a} and no experimental data exists that would allow access to them.

\begin{figure}[h!]
\begin{center}
    \includegraphics[width=0.36\textwidth]{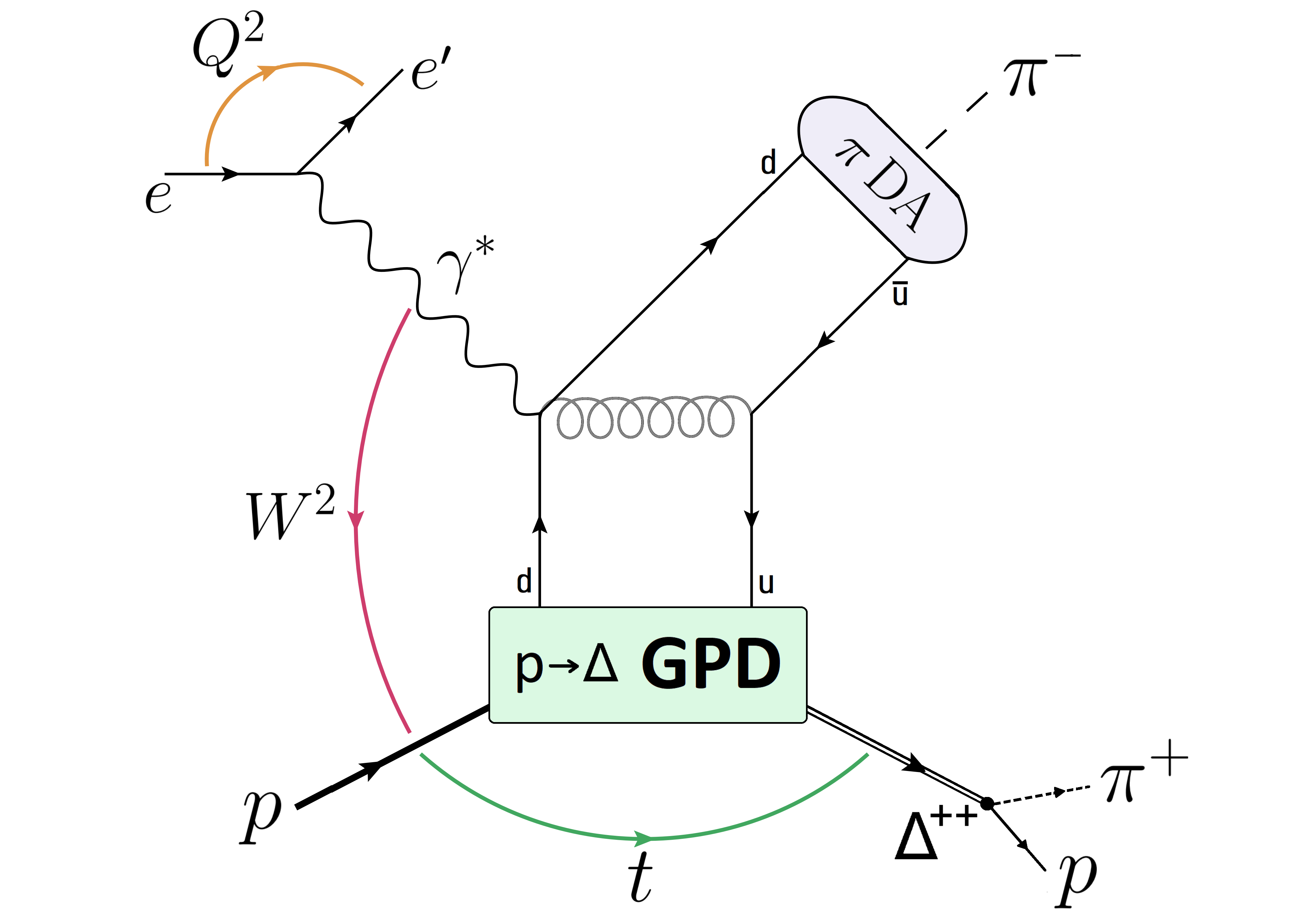} 
	\caption{Hard exclusive $\pi^- \Delta^{++}$ electroproduction off the proton in very forward kinematics ($-t/Q^2 \ll 1$) with the virtuality $Q^{2}$ and the four-momentum transfer $t$ to the $\Delta^{++}$.}
	\label{fig:production_mechanism}
	\end{center}
\end{figure}

The measurement of hard exclusive ($\gamma^{*}p \to \pi^{-}\Delta^{++} \to \pi^{-}[p\pi^{+}]$) electroproduction beam-spin asymmetries in this work is expected to represent a first observable sensitive to $N\to\Delta$ transversity transition GPDs, especially $G_{T_{5}}$ and $G_{T_{7}}$, and to $N\to\Delta$ transition GPDs in general. In analogy to ordinary GPDs \cite{Liutti}, it is expected that the production of charged pions is especially sensitive to the tensor charge of the $\Delta$ resonance, which is so far completely unexplored. 
As shown in Fig.~\ref{fig:production_mechanism}, the soft parts of the convolution can be described with transition GPDs and a meson distribution amplitude (DA).
It is assumed for this process that the QCD factorization theorem is valid within the Bjorken limit: $-t/Q^{2} \ll 1$ and $x_{B}$ fixed, with the additional condition $Q^2>>m^2_{\Delta}$ as discussed in Ref. \cite{13} for the $p \to \Delta$ deeply virtual Compton scattering (DVCS) process. However, a direct proof of the factorization for the investigated channel is not available yet.

Previous attempts to extract observables sensitive to $N\to N^{*}$ transition GPDs based on $p \to N^{*}$ DVCS ($e p \to e^{\prime} N^{*} \gamma$), but also based on $p \to N^{*}$ $\pi^{0}$ and $\pi^{+}$ production \cite{DHK81, SD22}, suffered from low statistics and a clean separation between the produced $\Delta$ events and the overlapping events from other nucleon resonances, as well as from non-resonant background. 
In contrast to these, the $p \to N^{*}$ $\pi^{-}$ production studied in this work with high statistics, focuses on the $\Delta^{++}$ resonance with an isospin $I_{z} = + 3/2$, which is only fulfilled by $\Delta$ resonances. Therefore, a large gap exists between the $\Delta(1232)$ and the higher-mass $\Delta$ resonances, starting at masses of 1.6~GeV and showing a strongly suppressed branching ratio compared to the $\Delta(1232)$, which allows a relatively clean extraction of the dominant $\Delta(1232)$ resonance and a clear identification and subtraction of the non-resonant background. Previous studies of this channel \cite{WDK78, DHK81} were strongly limited by low statistics and therefore constrained to the low $Q^{2}$ regime.

In exclusive electroproduction experiments, GPDs are typically accessed through differential cross sections and beam and target polarization asymmetries \cite{Dre1992, Are1997, Die2005}. The focus of this work is on the extraction of the structure function ratio  $\sigma_{LT'}/\sigma_{0}$ from beam-spin asymmetry (BSA) measurements.
In the one-photon exchange approximation the BSA is defined as \cite{Dre1992, Are1997}:
\begin{eqnarray}\label{eq:BSA}
	BSA  = \frac{\sqrt{2 \epsilon (1 - \epsilon)}  \frac{\sigma_{LT^{\prime}}}
	{\sigma_{0}}\sin\phi}
	{1 + \sqrt{2 \epsilon (1 + \epsilon)}\frac{\sigma_{LT} }{\sigma_{0}} \cos\phi
	+ \epsilon \frac{\sigma_{TT}}{\sigma_{0}}  \cos2\phi},
\end{eqnarray}
where the structure functions $\sigma_{L}$ and $\sigma_{T}$ that contribute to $\sigma_{0} = \sigma_{T} + \epsilon \sigma_{L}$, correspond to the contribution of longitudinally (L) and transversely (T) polarized virtual photons, and $\epsilon$ describes the polarization of the virtual photons. $\sigma_{LT}$, $\sigma_{TT}$ and the polarized structure function $\sigma_{LT^\prime}$ describe the interference between the amplitudes of longitudinally and transversely polarized virtual photons, with the prime in $LT^\prime$ indicating the dependence on the electron polarization. $\phi$ is the azimuthal angle between the electron scattering and the hadronic reaction plane. 

%%%%%%%%%%%%%%%%%%%%%%%%%%%%%%%%%%%%%%%%%%%%%%%%%%%%%%
%\section{\label{sec:Expsetup} Experimental Setup}

For the present study, hard exclusive $\pi^- \Delta^{++}$ electroproduction was measured at Jefferson Laboratory (JLab) with CLAS12 (CEBAF Large Acceptance Spectrometer for experiments at 12~GeV) \cite{VDB20}. The incident longitudinally polarized electron beam had energies of 10.2~GeV and 10.6~GeV, impinging on an unpolarized liquid-hydrogen target. The CLAS12 forward detector consists of six identical sectors within a toroidal magnetic field. The momentum and the charge of the particles were determined by 3 regions of drift chambers from the curvature of the particle trajectories in the magnetic field. The electron identification was based on a lead-scintillator electromagnetic sampling calorimeter in combination with a {\v C}herenkov counter. Pions and protons were identified by time-of-flight measurements.

%%%%%%%%%%%%%%%%%%%%%%%%%%%%%%%%%%%%%%%%%%%%%%%%%%%%%%
%\section{\label{sec:signal} Signal Extraction and Background Subtraction}

For the selection of deeply inelastic scattered electrons, cuts on $Q^{2} >$~ 1.5~GeV$^{2}$, the energy fraction of the beam carried by the virtual photon $y <$~0.75 and the invariant mass of the hadronic final state $W >$~2~GeV, were applied. To select the exclusive $e^{\prime} \pi^{-} \Delta^{++}$ final state, events with exactly one electron, one $\pi^{-}$ and one proton were detected, and the missing $\pi^{+}$ was selected via a cut on the $\pi^{+}$ peak in the $e'p\pi^{-}X$ missing mass spectrum and assigned to the missing 4-vector. %In addition, only forward kinematics were selected by a cut on $-t < 1.5$~GeV$^2$.
The dominant background from exclusive $\rho$ production was reduced by a cut on the invariant two pion mass $M_{\pi^+ \pi^-} > 1.1$~GeV to a level of less than 0.8\%. In addition, a cut on the $p\pi^{+}$ invariant mass $M_{p \pi^+} < 1.3$~GeV was applied for the final analysis to select the $\pi^-\Delta^{++}$ events and to reduce the non-resonant background that dominates below the tail of the resonance mass at larger $p\pi^{+}$ invariant masses (see also Refs. \cite{supl, SD22}). 

\begin{figure}[h!]
	\centering
		\includegraphics[width=0.45\textwidth]{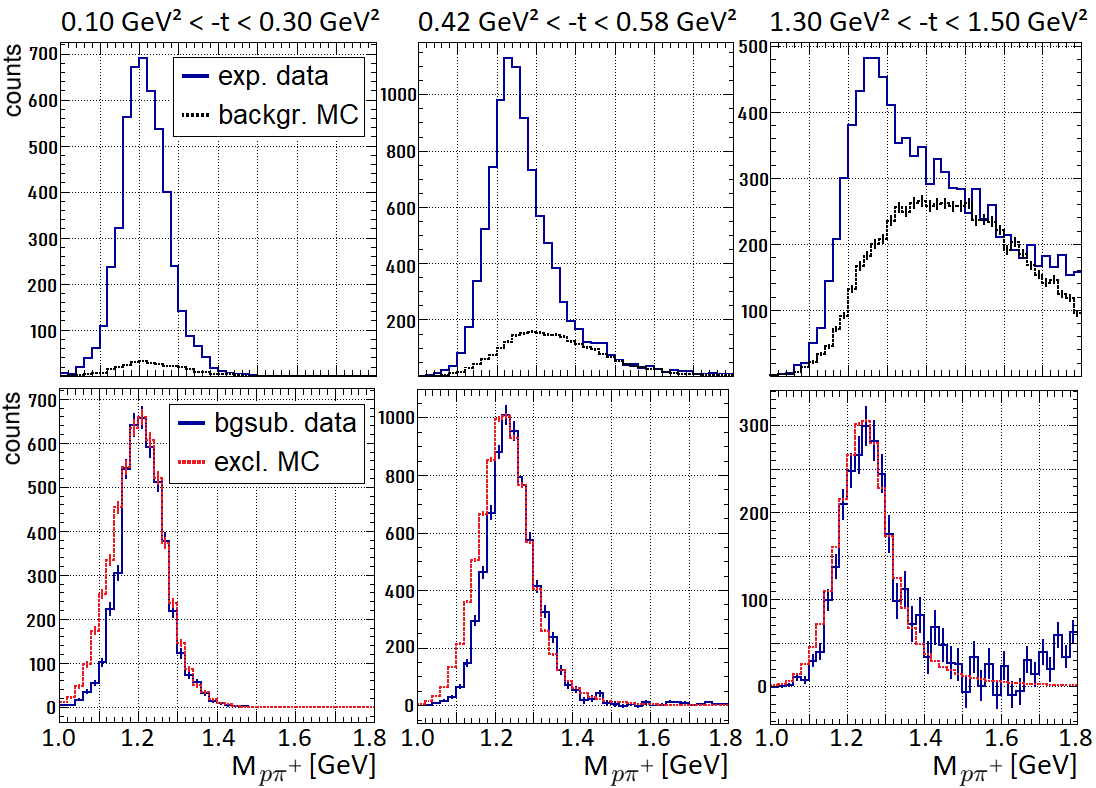}
	\caption{Upper row: $\Delta^{++}$ peak in the $p \pi^{+}$ invariant mass of the experimental data (blue, solid) in comparison to the non-resonant background obtained with the SIDIS MC (black, dashed) for selected bins of $-t$ in the forward region ($Q^{2}$ = 2.48~GeV$^{2}$, $x_{B}$ = 0.27) after a cut on $M_{\pi^+ \pi^-} > 1.1$~GeV. Lower row: $\Delta^{++}$ peak in the same bins after the subtraction of the background (blue, solid) in comparison to the result from the exclusive MC (red, dashed).}
	\label{fig:delta_mass}
\end{figure}

The remaining non-resonant background was studied by comparing the data to two Monte Carlo (MC) samples. The first sample was based on a full semi-inclusive deep-inelastic scattering (SIDIS) generator \cite{MSV92}, which contains all non-resonant background channels but not the exclusive $\pi^- \Delta^{++}$ production in forward kinematics. It can therefore be used to reproduce the background shape. As a second sample, an exclusive $\pi^- \Delta^{++}$ generator with literature values \cite{RLW22} for the mass and full width at half maximum of the $\Delta^{++}$ was used to reproduce the signal events. Both samples were processed through the full Geant4-based \cite{GEA03,Ung20} simulation and reconstruction chain. Good agreement for all underlying variables was observed.
Figure \ref{fig:delta_mass} (upper row) shows the $\Delta^{++}$ peak in the $p \pi^{+}$ invariant mass of the experimental data (without a cut on this mass) in comparison to the non-resonant background obtained with the SIDIS MC for selected bins of $-t$ in the forward region, integrated over $Q^{2}$ and $x_{B}$. Figure \ref{fig:delta_mass} (lower row) shows the $\Delta^{++}$ peak in the same bins after the subtraction of the background in comparison to the result from the exclusive MC. Both MC samples (signal + background) were scaled iteratively to match the measured distribution.
It can be observed that the non-resonant background is small close to $t_{min}$ but increases to $\approx$~40\% for the largest $-t$ bins considered, making a background subtraction necessary. The signal-to-background ratios were directly determined from the SIDIS MC in comparison to the experimental data. 

\begin{figure}[h!]
	\centering
		\includegraphics[width=0.43\textwidth]{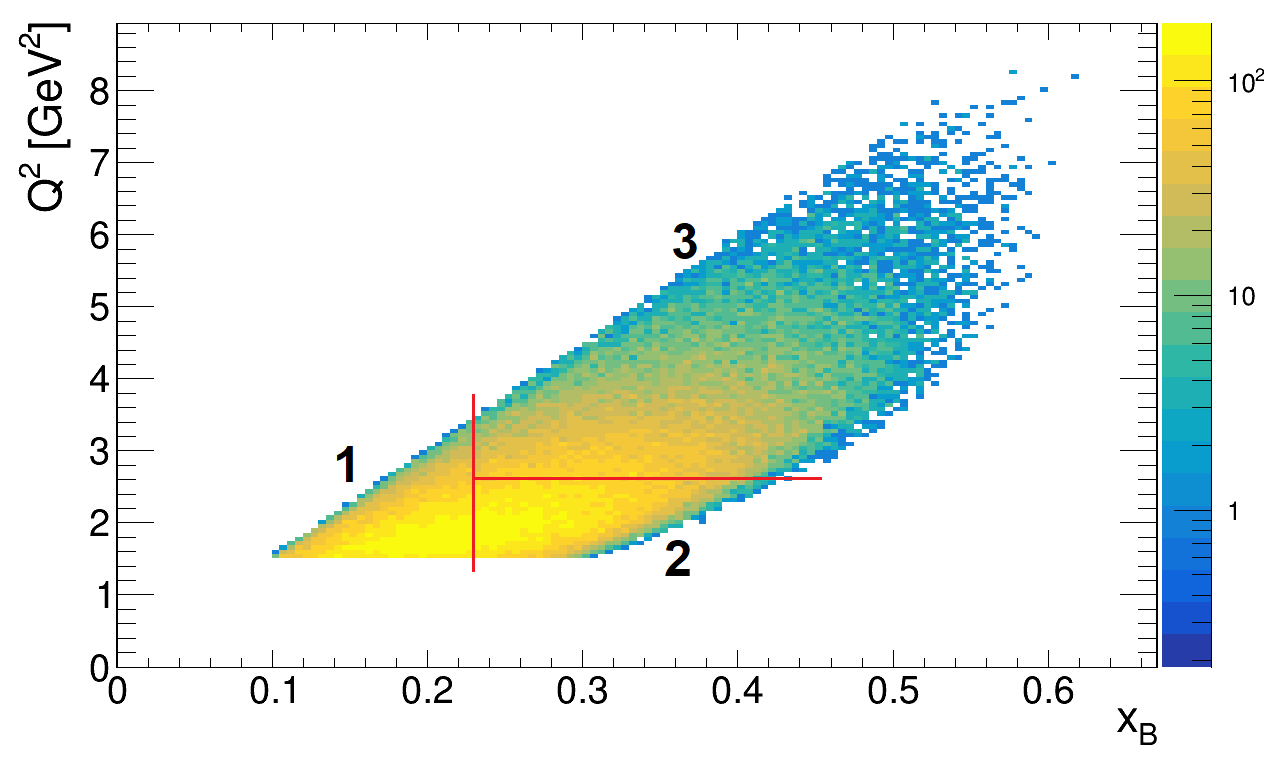}
	\caption{Distribution of $Q^{2}$ versus $x_{B}$ for $-t < 1.5$~GeV$^2$. The bin boundaries are shown as red lines (bin~1: $x_{B} <$~0.23; bin~2: $x_{B} >$~0.23, $Q^{2} <$~2.6~GeV$^{2}$; bin~3: $x_{B} >$~0.23, $Q^{2} >$~2.6~GeV$^{2}$).}
	\label{fig:q2x_bins}
\end{figure}

Figure \ref{fig:q2x_bins} shows the $Q^{2}$ versus $x_{B}$ distribution of the exclusive events, together with the applied binning scheme.
For each of the three $Q^{2}$-$x_{B}$ bins, up to seven bins in $-t$ and 9 bins in $\phi$ were defined to extract the BSA.
The BSA was determined experimentally from the number of counts with positive and negative helicity ($N^{\pm}_{i}$), in a specific bin $i$ as:
\begin{eqnarray}
	BSA_{i} = \frac{1}{P_{e}} \frac{N^{+}_{i} - N^{-}_{i}}{N^{+}_{i} + N^{-}_{i}},
\end{eqnarray}
\newline where $P_{e}$~=~86.6\%~$\pm$~2.7\% is the average magnitude of the beam polarization, which was measured with a M{\o}ller polarimeter upstream of CLAS12 \cite{VDB20}. 

The raw asymmetry was extracted from the defined signal region ($M_{p \pi^+} < 1.3$~GeV) and the background asymmetry, which was found to be between 0.0 and -0.04,
 was determined from the sideband in the region 1.45~GeV~$<$~$M_{p \pi^{+}}$~$<$~1.65~GeV. 
%It was checked that the non-resonant background from the neutral resonances ($N^{*} \to p\pi^-$) homogeneously spreads over the sideband and signal region in the $p \pi^{+}$ invariant mass. 
The signal asymmetry was extracted by subtracting the background asymmetry from the raw asymmetry weighted with the signal to background ratio in each $Q^{2}$, $x_{B}$ and $-t$ bin.

%As a second completely independent crosscheck, a bin-by-bin background subtraction was performed based on a fit of the complete distribution (signal + background) with a so-called ``Sill'' function \cite{16} plus a fifth-order polynomial background in each $Q^{2}$, $x_{B}$, -$t$ and $\phi$ bin and for each helicity state. After the combined fit, the signal and background contributions were separated and the asymmetry was calculated based on the pure signal events. It was found that both methods provided consistent results for the signal asymmetry within the statistical uncertainty.

\begin{figure}[h!]
	\centering
		\includegraphics[width=0.48\textwidth]{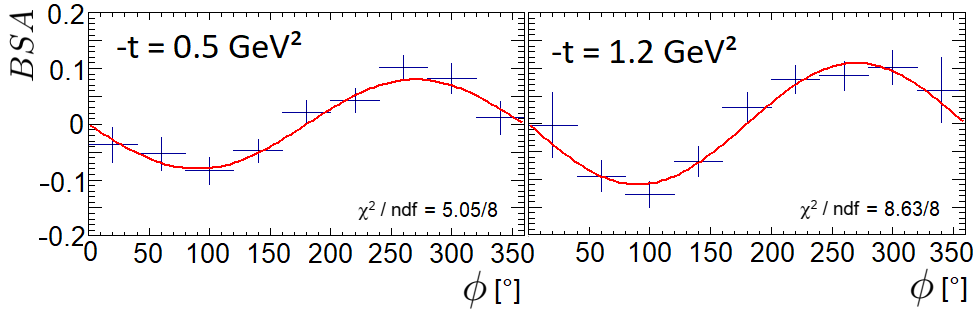}
	\caption{BSA as a function of $\phi$ for representative $-t$ bins ($Q^{2}$ = 2.48~GeV$^{2}$, $x_{B}$ = 0.27). The red line shows the $\sin\phi$ fit.}
	\label{fig:sinphi_fit}
\end{figure}

To extract the structure function ratio $\sigma_{LT'}/\sigma_{0}$, the dependence of the BSA on the azimuthal angle $\phi$ (see Fig. \ref{fig:sinphi_fit}) was fit to Eq. \ref{eq:BSA}.
It can be seen that a precise measurement of the $\phi$ dependence, which can be well described by a $\sin\phi$ shape, is possible. 

The main source of systematic uncertainty is given by the background subtraction. It was determined by varying the signal-to-background ratio and the background asymmetry within the estimated uncertainty ranges and was found to be on the order of 1.5 - 12.5\% (depending on the $-t$ bin). 
%Also the difference between the sideband-based background subtraction and the bin-by-bin-based background subtraction was found to be within these limits. 
Also the impact of the denominator terms in Eq. (\ref{eq:BSA}) on $\sigma_{LT'}/\sigma_{0}$ was evaluated and found to be on the order of 2.8\%, which was treated as part of the systematic uncertainty. The systematic effect due to the uncertainty of the beam polarization (3.1\%) was determined based on the uncertainty of the measurement with the M{\o}ller polarimeter. A Geant4-based MC simulation \cite{Ung20} was performed to estimate the impact of acceptance and bin-migration effects (2.9\%). Also acceptance effects from the decay products of the $\Delta^{++}$ were evaluated and found to be of the same order. Systematic uncertainties due to radiative effects (3.0\%) have been studied based on Ref. \cite{AAB02}. Several additional sources of systematic uncertainty, including particle identification and the effect of fiducial volume definitions, were found to be small ($<$2.0\%). The total systematic uncertainty in each bin was defined as the square-root of the quadratic sum of the uncertainties from all sources. On average it was found to be on the order of 7.1 - 14.3\% (depending on the $-t$ bin), which is smaller than the statistical uncertainty in most kinematic bins.

%%%%%%%%%%%%%%%%%%%%%%%%%%%%%%%%%%%%%%%%%%%%%%%%%%%%%%
%\section{\label{sec:results} Result and Discussion}

\begin{figure*}
	\centering
		\includegraphics[width=0.95\textwidth]{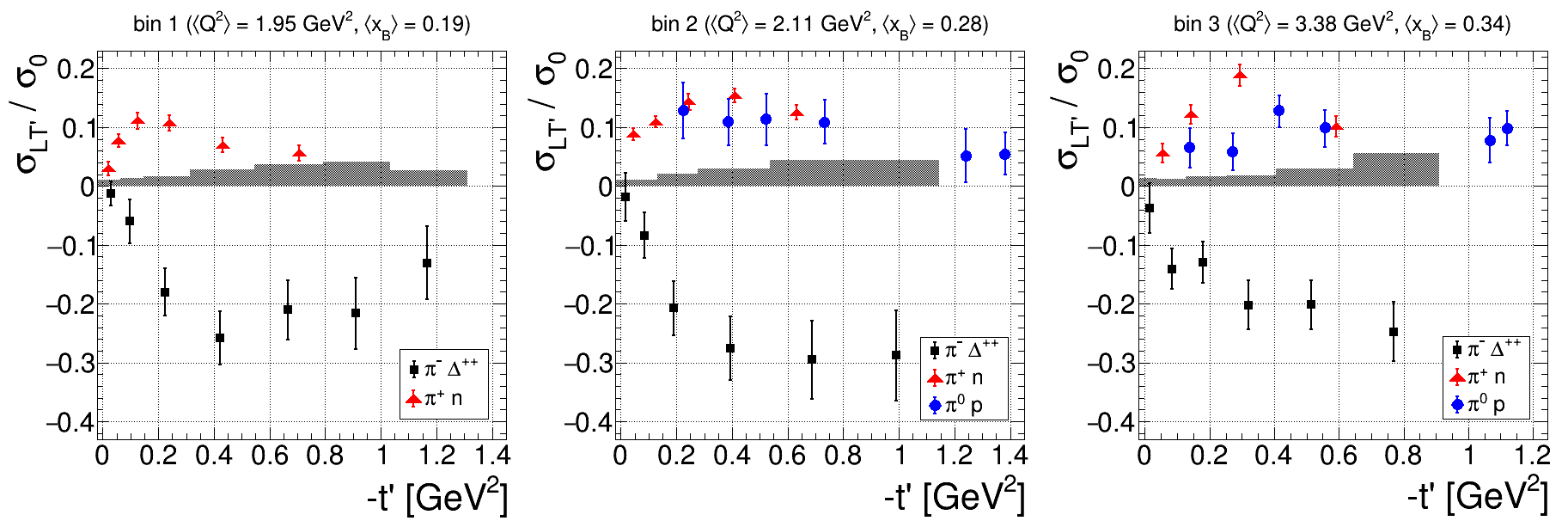}
	\caption{$\sigma_{LT'}/\sigma_{0}$ and its statistical uncertainty for $\pi^- \Delta^{++}$ (black squares, this work) as a function of $-t'= (|t|-|t_{min}|)$ in the forward kinematic regime and its systematic uncertainty (gray band). The sub-figures correspond to the results for the different $Q^{2}$ and $x_{B}$ bins defined in Fig. \ref{fig:q2x_bins}. The mean kinematics are shown on top of each sub-figure. The corresponding result tables can be found in the supplemental material \cite{supl} and can be downloaded from Ref. \cite{CLASdata}. For comparison, the results from the hard exclusive $\pi^+ n$ (red triangles, Ref. \cite{Diehl23}) and $\pi^0 p$ (blue circles, Ref. \cite{Kim23}) electroproduction with similar kinematics are shown.}
	\label{fig:result_final}
\end{figure*}

Figure \ref{fig:result_final} shows the final results for $\sigma_{LT'}/\sigma_{0}$ in the region of small $-t$, where a description based on transition GPDs is expected to be applicable, and compares them to measurements from the hard exclusive $\pi^+ n$ and $\pi^0 p$ electroproduction from Refs. \cite{Diehl23, Kim23}, which can be described with ground state GPDs. The result tables for $\pi^- \Delta^{++}$ can be found in the supplemental material \cite{supl}.
The structure function ratio $\sigma_{LT'}/\sigma_{0}$ for $\pi^- \Delta^{++}$ is clearly negative in all kinematic bins and shows a shape that can be explained by the contributing structure functions. The integrated cross section $\sigma_{0} = \sigma_{T} + \epsilon \sigma_{L}$, which provides the denominator of the ratio, is typically forward peaked due to the pion pole term contribution, while $\sigma_{LT^\prime}$ which is the numerator of the measured ratio, is constrained to be zero at $t=t_{min}$ due to angular momentum conservation \cite{previous1, 13a}. This behavior can be observed for $\pi^{-}$ as well as for $\pi^{+}$, while $\pi^{0}$ shows a more constant behavior with a continuous decrease for increasing $-t$ over all kinematic bins due to the missing pion pole term.

As an interesting feature, the magnitude of the structure function ratio for $\pi^{-}$ production is approximately two times larger than for $\pi^{+}$ production and has an opposite sign. It is known, that the pion pole contribution does not change the sign of the structure function ratio and that it has approximately the same magnitude for $\pi^{+}$ and $\pi^{-}$ with an uncertainty of 10 \% in the measured $Q^{2}$ regime \cite{PK23}.
The opposite sign can be directly explained by the quark polarization. For $e p \to e^{\prime} n \pi^{+}$, the polarized $\gamma^{*}$ removes a longitudinally polarized $u$-quark from the proton ($uud$). This $u$ quark then combined with a $\bar{d}$ from a $d\bar{d}$ vacuum pair with the $d$-quark returning to create the final state neutron ($udd$). In contrast to this, for $e p \to e^{\prime} \Delta^{++} \pi^{-}$, the polarized $\gamma^{*}$ kicks out a longitudinally polarized $d$-quark from the proton, which combines with a $\bar{u}$ from a $u\bar{u}$ vacuum pair with the $u$-quark returning to produce the final state $\Delta^{++}$ ($uuu$). For the $\pi^{0}$ a clear assignment to one quark type cannot be made due to the mixed content of its wave function.
It is known that within the valence region, the polarization of $d$-quarks, $\Delta d$, in the proton is negative, while the polarization of $u$-quarks, $\Delta u$, is positive and that the polarization of $\bar{u}$- and $\bar{d}$-quarks in the proton is small \cite{CMMS22}.
If we now look into the $\pi^{+}$ ($\left|u\bar{d}\right\rangle$) and $\pi^{0}$ ($1/\sqrt{2} \left[\left|u\bar{u}\right\rangle - \left|d\bar{d}\right\rangle\right]$) BSAs, both are positive. The $\pi^{+}$ production is clearly dominated by $u$-quarks, while for $\pi^{0}$ the negative sign in front of the $d \bar{d}$ part of the wave function turns the polarization contribution from the $d$-quark around and causes a sizable positive asymmetry. This asymmetry is in some regions similar to $\pi^{+}$ even though there is no amplification by the pion pole. On the other hand, $\pi^{-}$ ($\left|d\bar{u}\right\rangle$) BSAs are negative, since they are dominated by $d$-quarks. The reaction $e p \to e^{\prime} \Delta^{++} \pi^{-}$ can therefore provide access to the polarized $d$-quark content within the proton, which is otherwise hard to probe. Since the described effects are coming from polarized quarks, even bigger effects with asymmetries on the order of 40\% \cite{13a} are expected for double spin asymmetries with a longitudinally polarized beam and a longitudinally polarized target.

The absolute magnitude of the $u$- and $d$-quark polarization is similar in the proton \cite{CMMS22}. Therefore, a similar absolute magnitude of the BSA would be expected for $\pi^{+}$ and $\pi^{-}$ based on these simple considerations. To correctly model the magnitude of the asymmetry, the excitation process from the ground state proton to the $\Delta^{++}$ resonance has to be considered through transition GPDs. According to this formalism, the measured cross section ratio $\sigma_{LT'}/\sigma_{0}$ is expected to be a twist-3 quantity ($\sim 1/Q$). The comparison of $\sigma_{LT'}/\sigma_{0}$ in bin 2 (low $Q^{2}$, high $x_{B}$) and bin 3 (high $Q^{2}$, high $x_{B}$) shows in average a slight decrease of the magnitude, which would be consistent with this assumption, but not conclusive within the uncertainties of our measurement.
In Ref. \cite{13a} the first transition GPD-based predictions for the unpolarized partial cross sections of the $e p \to e^{\prime} \Delta^{++} \pi^{-}$ process have been made. In the large $N_{C}$ limit it is expected that the process is dominated by the transversity transition GPDs $G^{3}_{T_{5}}$ and $G^{3}_{T_{7}}$ (the superscript indicates the twist-3 nature), which can be related to the ground state transversity GPD $H_{T}$ \cite{13a}.
Therefore, it can be assumed that the polarized structure function $\sigma_{LT'}$, which is given by products of convolutions of transversity and helicity non-flip transition GPDs with sub-process amplitudes, shows the following relation to the two dominant transversity transition GPDs:
\begin{equation}
	\sigma_{LT'} \sim \sqrt{-t'}~~Im\left[ G^{3}_{T_{5}} \cdot A +  c~G^{3}_{T_{7}} \cdot A' \right],
\end{equation}
with an unknown kinematic factor $c$ and helicity amplitudes for longitudinally polarized virtual photons $A$ and $A'$, which are determined by the helicity non-flip transition GPDs $\widetilde{G}_3$ and $\widetilde{G}_4$ within the large $N_{C}$ limit.

Due to the large uncertainties of the so far nearly unconstrained transition GPDs, the existing predictions on unpolarized cross sections have large uncertainties \cite{13a}. For asymmetries, these uncertainties are expected to be even larger due to the dependence on the imaginary part of helicity amplitude products and the related, so far poorly known, relative phases of the helicity amplitudes, making reliable predictions at the present stage impossible \cite{13a}.
So far, only transition GPD-based predictions for the BSA of the $p \to \Delta$ DVCS process, based on the twist-2 transition GPDs, exist \cite{13}. For this process, the BSA of the $\Delta$ production is excepted to be approximately 20-40\% larger than for the regular DVCS process of the ground state proton \cite{13}. This difference can be directly related to the increase of the magnitude of the underlying twist-2 transition GPDs. Assuming that the twist-3 transition GPDs are affected from the inelasticity in a similar way, this would lead to an increase by a factor 1.2-1.4 for the BSA of $\pi^{-}$ in comparison to $\pi^{+}$.
However, based on these considerations the observed effect can not be completely explained. More theoretical investigations and especially more experimental constraints are necessary to obtain a reliable parameterization of the transition GPDs and a reliable description of the hard exclusive $N^{*}\pi$ production process.

%%%%%%%%%%%%%%%%%%%%%%%%%%%%%%%%%%%%%%%%%%%%%%%%%%%%%%
%\section{\label{sec:summary} Summary}

In summary, we have performed a first multidimensional measurement of the structure function ratio $\sigma_{LT'}/\sigma_{0}$ for $\vec{e} p \to e^{\prime} \pi^- \Delta^{++}$ at large photon virtualities above the resonance region. The results have been discussed in the context of quark polarizations and in relation to $p \to \Delta$ transition GPDs. The measurement can give us a direct access to the $d$-quark content of the nucleon and can be seen as a first measured observable sensitive to $p\to\Delta$ transition GPDs. The observed results in comparison to the $\pi^+ n$ and $\pi^0 p$ final state, agree well with the expectations for the effects of the inelasticity introduced to the GPDs for the $p\to\Delta$ transition. 
The measurements presented in this work have initiated first theoretical investigations of the hard exclusive $\pi^- \Delta^{++}$ production based on transition GPDs \cite{13a}. This opens the path to the investigation of the 3D structure of resonances from future measurements of the $N \to N^{*}$ DVCS process, as well as other $N \to N^{*}$ deeply virtual meson production (DVMP) channels at JLab and at the future EIC with an extension to the strangeness sector.

%%%%%%%%%%%%%%%%%%%%%%%%%%%%%%%%%%%%%%%%%%%%%%%%%%%%%%%%%%%%%%%%%%%%%%%%%%%%%%%%%%%%%%%%%%%%
%%%%%%%%%%%%%%%%%%%%%%%%%%%%%%%%%%%%%%%%%%%%%%%%%%%%%%%%%%%%%%%%%%%%%%%%%%%%%%%%%%%%%%%%%%%%

We acknowledge the outstanding efforts of the staff of the Accelerator and the Physics Divisions at Jefferson Lab in making this experiment possible. 
We owe much gratitude to P. Kroll for many fruitful discussions concerning the interpretation of our results.
This work was supported in part by the U.S. Department of Energy, the National Science Foundation (NSF), the Italian Istituto Nazionale di Fisica Nucleare (INFN), the French Centre National de la Recherche Scientifique (CNRS), the French Commissariat pour l$^{\prime}$Energie Atomique, the UK Science and Technology Facilities Council, the National Research Foundation (NRF) of Korea, the Helmholtz-Forschungsakademie Hessen f\"ur FAIR (HFHF), the Deutsche Forschungsgemeinschaft (DFG) and the Chilean Agency of Research and Development (ANID). The Southeastern Universities Research Association (SURA) operates the Thomas Jefferson National Accelerator Facility for the U.S. Department of Energy under Contract No. DE-AC05-06OR23177.

%%%%%%%%%%%%%%%%%%%%%%%%%%%%%%%%%%%%%%%%%%%%%%%%%%%%%%%%%%%%%%%%%%%%%%%%%%%%%%%%%%%%%%%%%%%%
%%%%%%%%%%%%%%%%%%%%%%%%%%%%%%%%%%%%%%%%%%%%%%%%%%%%%%%%%%%%%%%%%%%%%%%%%%%%%%%%%%%%%%%%%%%%

\end{document}